\documentstyle[11pt,aaspp4]{article}

\lefthead{Ram\'{\i}rez, de Souza \& Schade}
\righthead{Shape of Orbits in CNOC1 Clusters}
     
\def\kms{\ifmmode{\,\hbox{km}\,s^{-1}}\else {\rm\,km\,s$^{-1}$}\fi}

\def\Mpc{\ifmmode{h^{-1}\,\hbox{Mpc }}\else{$h^{-1}$\thinspace Mpc }\fi}

\begin{document}

\title{Shape of the Galactic Orbits in the CNOC1 Clusters}

\author{Amelia C.  Ram\'{\i}rez}
\affil{University  of  S\~ao  Paulo, Instituto Astron\^omico e Geof\'{\i}sico,
Astronomy  Department}
\authoraddr{Av. Miguel Stefano 4200, Agua Funda, CEP:04301-904, 
S\~ao Paulo, Brazil}
\authoremail{aramirez@iagusp.usp.br}

\author{Ronaldo  E.  de Souza }
\affil{University  of  S\~ao  Paulo, Instituto Astron\^omico e Geof\'{\i}sico,
Astronomy  Department \\ 
and \\
University of Arizona, Steward Observatory}
\authoraddr{Av. Miguel Stefano 4200, Agua Funda, 
CEP:04301-904, Sao Paulo, Brazil}
\authoremail{ronaldo@as.arizona.edu}
\author{David Schade}
\affil{Dominion Astrophysics Observatory}
\authoraddr{5071 West Saonich Road, Victoria B.C. V8X4M6, Canada}
\authoremail{schade@dao.nrc.ca}

\slugcomment{Accepted in the Astrophysical Journal, April 2000}

\begin{abstract}

We present an analysis of the orbital properties in 9
intermediate-redshifts cluster of the CNOC1 survey and we compare them
to a control sample of 12 nearby clusters.  Similar to the nearby
elliptical galaxies, the bulge-dominated galaxies in clusters at
redshifts $\sim 0.1-0.4$ present orbits that are more eccentric than
those for disk-dominated galaxies. However, the orbital segregation is
less significant than that found for elliptical and spiral galaxies
in nearby cluster.  When galaxies are separated by colors | red
galaxies with colors in the rest frame (U-V)$_o > 1.4$, and blue
galaxies with (U-V)$_o \leq 1.4$ | the strongest orbital segregation is
found. Therefore, the segregation we found seems to modify more
efficiently the star formation activity than the internal shape of the
galaxies.  When we compare the orbits of early-type galaxies at
intermediate-redshift with those for z=0, they seem to develop
significant changes getting much more eccentric.  A different behavior
is observed in the late-type galaxies, which present no-significant
evolution in their orbit shapes.

\end{abstract}


%

\section{Introduction}

A successful model of formation and evolution of galaxy clusters must
explain, among other properties, the observed morphology segregation
found in nearby rich clusters.  The first detected difference between
galaxy populations was the strong gradient in morphological type with
radial distance, producing a concentrated spatial distribution of
early-type in comparison with the sparse distribution of late-type
galaxies.  This effect was associated to the well discussed
morphology-density relation, $T-\Sigma$ (Dressler 1980), or to the
alternative explanation of morphology-radius relation $T-R$ (Sanrom\`a
\& Salvador-Sol\'e 1990, Whitmore, Gilmore \& Jones 1993). These two
relations allow to attribute the morphology of galaxies to local or to
global cluster properties, respectively.  The T-$\Sigma$ relation is
observed in both open and compact clusters in the local universe.
However, this is apparently not the case at intermediate-redshift.
Dressler et al. (1997) find that the T-$\Sigma$ relation is
qualitatively similar in compact clusters at intermediate-redshifts,
but completely absent in the open clusters at a similar epoch.  This
result suggest that morphological segregation occurs hierarchically
over time, i.e.  groups that make up the irregular, open clusters at
intermediate-redshifts have not undergone significant morphological
segregation to establish a T-$\Sigma$ relation. However, by the
present epoch, the groups that make up the local open clusters would
have had sufficient time to establish such correlation.  If this
hypothesis is correct, clusters at high redshifts should show little
or no morphological segregation.

Another segregation in nearby clusters, reported by many authors (Moss
and Dickens 1977, Tully \& Shaya 1984, Huchra 1985, Dressler \&
Shectman 1988, Sodr\'e et al.  1989, Bird et al.  1994, Andreon 1994,
Biviano et al.  1997, Colless \& Dunn 1996, Fadda et al.  1996,
Girardi et al.  1996, Scodeggio et al.  1995, Andreon \& Davoust 1997
and Andreon 1996), is identified as a kinematical segregation,
characterized by a early-type population with a velocity dispersion
always lower than the velocity dispersion of the late-type population.
That effect is commonly explained as a consequence of the different
virialization state of both populations, where the observed behavior
of late-type is attributed to the possibility that they have been
accreted by the cluster more recently, after the collapse and violent
relaxation of the initial population of early-type galaxies which now
constitute the cluster core.  We can also mention, continuing with the
detected kinematical segregation in clusters, the less studied
luminosity segregation (Biviano et al. 1992, Fusco-Fermiano \& Menci
1998, Kashikawa et al. 1998), which seems to affect preferentially the
more luminous early-type galaxies. Biviano et al. (1992) found that
this type of galaxies are located in the center of clusters and they
have velocities with respect to the mean cluster velocity lower than
other less massive members, as could be expected if they are affected
by dynamical friction. Recently, another type of galaxy segregation
was detected by Ram\'{\i}rez and de Souza (1998, hereafter RdS98).
They analyzed kinematically the early- and late-type galaxies in 18
nearby rich clusters, and they concluded that inside 1.0 \Mpc
elliptical galaxies present orbits more eccentric than spiral
galaxies, and only well outside the fudicial radius r$_{200}$ the
orbits of spiral galaxies become more radial.  A possible explanation
for it is related to the stability of the morphological shapes of
galaxies as they plunge towards the central regions.  In this case
objects with roughly circular orbits, will not have their morphologies
seriously affected because they avoid the cluster center where the
probability of occurring a strong interaction will be higher.
However, those with more eccentric orbits will cross the densest
cluster regions and will experience on average a stronger
environmental influence and a higher encounter rate.  Furthermore, if
along the cluster life-time there are significant orbital changes of
their members, high redshifts clusters could show a different level of
orbital-segregation than the nearby population.

In this paper we present a kinematical analysis of clusters at
intermediate redshifts. The goal of this analysis is to detect how
strong the orbital-segregation between early- and late-type galaxies
at this epoch is.  The results are compared with those for a nearby
clusters sample.  In section 2 we present the distribution function we
used to analyze the line-of-sight velocities of a cluster having a
velocity field with anisotropy.  In section 3, it is defined two sample of
clusters and we define the morphology classification we
use.  In section 4 we study the morphological segregation in the
intermediate-redshift sample. In section 5 we use the average
deviation of the line-of-sight velocity normalized to the velocity
dispersion to trace the orbit distributions of early- and late-type
galaxies.  In section 6 we summarize our main conclusions.

\section{Distribution of velocities and the average deviation}

In this section we present a brief summary of the analytical analysis,
completely presented in RdS98, of the velocity distribution of systems
with velocity anisotropies.

Let us assume that for a given morphological class the velocity
distribution function is Gaussian, having however different
dispersions along the radial ($\sigma_R$) and transversal directions
($\sigma_\perp$).  The behavior of the velocity distribution of this
system can therefore be described by the anisotropy parameter $\eta =
\sigma_R/\sigma_\perp$.  A large value of $\eta$ for a given
population means that its members are crossing the cluster with an
almost radial orbit, and therefore are more sensitive to suffer
gravitational encounters with objects in the dense central regions.
On the contrary, objects with lower anisotropy parameter tend to have
a more circular orbit with small penetration in the dense core region.
Therefore, for a given morphological class the anisotropy parameter
should allow us to connect the efficiency of the environment
perturbations with the related kinematical orbital behavior.

The line-of-sight velocity distribution in a Gaussian velocity field
with a fixed anisotropy can be expressed as:

\begin{equation} F(u;\eta ) = \frac{1}{(2\pi)^{1/2}}
			  \int^1_0        \frac{1}{\Theta(\omega,\eta)}
			 e^{-\frac{u^2}{2\Theta(\omega,\eta)^2}}
			 d\omega
\end{equation}

\noindent where $\omega = \cos \theta$, $u = v_z/\sigma$ and $v_z$ is
directed along the line-of-sight. The velocity dispersion $\sigma$ can
be expressed in the form $\sigma = \sqrt{(\sigma_R^2 +
2\sigma_\perp^2)/3}$ and the term $\Theta(\omega,\eta)=
\sqrt{\frac{3(1-\omega^2 +\eta^2\omega^2)}{2+\eta^2}}$ represents a
correction of the velocity dispersion due to the presence of the
anisotropic field.  In particular, for $\eta=1$ we retrieve the
expected Gaussian shape for an isotropic velocity field.  Then, even
if the radial and transversal distributions are assumed to be
Gaussian, the observed distribution along the line-of-sight, when the
anisotropy parameter is different from one, is not Gaussian.

Although in the general case the line-of-sight velocity density
distribution can be estimated only by numerical methods, it is
interesting to note that the expressions for the moments can be solved
exactly.  The distribution $F(u;\eta)$ is symmetric by construction,
resulting that the first centered moment is obviously zero.  The
second moment corresponds to the variance of the distribution that
remains independent of the anisotropic parameter.  Therefore, we may
expect that two populations responding to the same gravitational 
potential can
present different orbital shapes, but their velocity dispersions
remain constant. Then, to detect anisotropies is better to work
with those expectation values that are function of the $\eta$
parameter. The more interesting of them is the average or mean
deviation (Kendall, Stuart and Ord, 1987) of the line-of-sight
velocity normalized to the velocity dispersion

\begin{equation} |u| = \sum_{i=1}^N |(v_z)_i|/\sigma 
\end{equation}

\noindent The predicted value of this statistical
parameter, using $F(u;\eta)$, can be estimated by the expressions,


for $\eta < 1$

\[ |u| = \sqrt{\frac{6}{\pi (\eta^2+2)}} \biggl( \frac{\eta}{2} + 
\frac{1}{2}  \frac{1}{   \sqrt{1-\eta^2}}  sin^{-1}(\sqrt{1-\eta^2}  )
\biggr) \]

and for $\eta > 1$

\[ |u| = \sqrt{\frac{6}{\pi (\eta^2+2)}} \biggl( \frac{\eta}{2} + 
\frac{1}{2} \frac{1}{ \sqrt{\eta^2 - 1}} ln(\sqrt{\eta^2 - 1} + \eta )
\biggr), \]

\noindent the behavior of this function is presented in
 Figure 1 and the limits it reaches are the following:

\vspace{0.5cm}

(i) population with  radial orbits  

~~($\eta \gg 1) \longrightarrow  |u| \simeq\sqrt{3/2\pi} \simeq 0.69$

\vspace{0.3cm}

(ii) population with purely  circular orbits  

~~($\eta  = 0) \longrightarrow  |u|  =\sqrt{3\pi }/4 \simeq 0.77$ 

\vspace{0.3cm}

(iii) population with isotropic orbits 

~~($\eta =  1) \longrightarrow  |u| = \sqrt{2/\pi}\simeq 0.80$

\vspace{0.5cm}

From Figure 1 we can observe that the average deviation as a function
of the anisotropic parameter $\eta$ has a maximum at $\eta=1$. The
function is bi-valuated between $|u|$= 0.77 to 0.80, producing an
indetermination when we try to derive the value of $\eta$.  As a
consequence we cannot distinguish between a purely circular model
($\eta = 0$) and one having a radial contribution slightly higher than
the isotropic case ($\eta \leq 2.4$).  However, the region of 
$|u| < 0.77$ or $\eta > 2.4$ is free from this problem, and as a
consequence the models of highly radial orbits can be easily
discriminated. Therefore, the average deviation is easy to 
measured in real clusters, using the equation (2), and a direct comparison 
with the expected values for each orbit family could be done.

\subsection{Comparison with other methods}

We must remark that our simplified treatment of the velocity
distribution produces the degeneracy in the average deviation value when
it is a function of the anisotropy parameter.  
This is a consequence of
two assumptions we made, the first is the velocity
distribution as a Gaussian with constant anisotropy, and the second
is related to the density distribution which is only restricted to a
spherical symmetrical distribution.  In fact, the distribution function 
is not necessarily unique because the 
density distribution, $\rho(\psi,r)$, could be originated from many 
gravitational potential.

A usual method to fix the degeneracy is by the use of a complete set
of defined functions which follow the Jeans's equations and the Poisson's
equations in a self-consistent way. The unknown functions which
well define the system | i.e. spatial density distribution, 
velocity dispersion profiles along the radial and tangential direction, 
and the mass profile | could be obtained by fitting and inverting 
the observed functions, i.e. projected velocity dispersion profiles
 and surface brightness profile.
It is usually assumed a given profile for the anisotropy 
together with a  M/L distribution. In fact, 
most of the works dealing with
these self-consistent distribution functions  assume that M/L is constant
and the anisotropy is a function of the radius alone 
(e.g. Binney \& Mamon 1982, The \& White 1986, Merritt 1987).
However, the solutions given by inverse problem are limited to systems with 
high number of measured velocities | i.e. few 
clusters or synthetic clusters produced by adding others | 
since the errors or incompleteness in the
data will be amplified when going from data space to model space.
Then, in most of the cases the results tend to be noisy,
unless some objective smoothness condition is placed on the
solution. An approach is to replace the data by ad hoc 
fitting functions for which the inversions can be carried out exactly, 
or to use smooth functions that are generated from the data using
nonparametric algorithm (e.g. Qian et al. 1995). 
For axisymmetric system Merritt (1996) concluded that
they are able to construct the gravitational potential and the kinematic 
 by the use of the full two-dimensional set of velocity, 
 line-of-sight velocity distribution (LOSVD) along
 the radius. Binney et al. (1990) pioneered this method using only few 
cuts along the radius. Both methods are very useful 
when we have enough resolution in the shape of the LOSVD at different radius, 
because  it depends on the high order moments of the distribution. The
nonparametric fits of the LOSVD also enables  
to detect the variation of the anisotropy without making  
any hypotheses about the ratio M/L, although it is necessary to guess an 
initial form of the gravitational potential. 

When we compare our method with the models cited above, 
we must remember that we are assuming
the anisotropy constant along the radius. 
This difference produces that the
variability of the LOSVD along the radius is not due to a variability 
in the anisotropy, instead it must be related to the mass
distribution and how well the galaxies trace the mass.
Furthermore, we must have in mind that we are 
studing the global average deviation, which represents the
behavior of the integrated average deviation along the projected radius, 
and we are integrating along the inner region of the cluster together 
with a part of the outer region.

Nevertheless, if we want to get an idea of the behavior of the average 
deviation along the radius, we can compare our model to these of 
variable anisotropy as for example the models of Merritt (1987),
Dejonghe (1987), Gerhard (1993) and Merritt \& Gebhardt (1996). 
They used different density profiles
and different gravitational potential to obtain that almost always in
the tangentially-anisotropic systems, i.e. $\sim \eta=0$, the LOSVD 
tends to have a peaked shape in the center and evolves to a flat-topped
shape in the outer regions.
In the radial-anisotropic systems, $\eta > 2.4$, they have the LOSVD 
shapes slightly flat-topped in the center an evolve to more peaked 
shapes in the outer regions. These behavior must produce very different
variation  in the average deviation along the radius, specially in the 
outer region, and these variations must be stronger in the 
tangentially-anisotropic systems  than in the radial-anisotropic systems. 
 The center and the outer radius 
in these cases are scaled by the core radius and they represent  
a very central and a very external region within a galaxy cluster. 

On other hand, if we compare our simple model with  models 
that assume the function distribution of quasi-separable form 
(Gerhard 1991) and these ones that expand the distribution 
function in the standard Hermite polynomials (Gerhard 1993),
we are approximately restricted to the terms zero-order Gaussian 
and first-order odd  of such distribution functions. 
The goodness of the simplicity of our method is achieved 
at the cost of having to determine not exactly which family of 
orbits are more common when the studied systems have circular, 
isotropic or mildly radial orbits, however, the highly radial family 
are easily identified. This characteristic enable us to detect segregation 
in the velocity anisotropy of our sample, because early and late-type 
galaxy populations  have anisotropies that are not constrained to the 
degenerated region of the average deviation curve.

Also, we must remark  that we are using the average 
deviation to get the anisotropy of the system, which  corresponds  
to a first order moment of the distribution. This is  
more robust than higher order moments and it  was always 
disconsidering in the past mainly due to the difficulty of
dealing with it analytically.

\section{The sample selection: intermediate-redshifts and nearby clusters}

Our basic sample contains all clusters observed by the Canadian
Network for Observational Cosmology, CNOC1 (Carlberg et al 1996). It
contains $\sim$2600 galaxies having Gunn $r$ magnitudes, $g-r$ colors,
and redshifts.  They were selected in the fields of 16 high luminosity
X-ray clusters spread from redshift 0.18 to 0.55.  This is a
relatively homogeneous sample of clusters that is guaranteed to be at
least partially virialized on the basis of their X-ray emission.
Galaxies of all colors above a k-corrected Gunn $r$ absolute magnitude
$M_r^k$ of $-18.5$ are considered in the calculations.  Together with
the CNOC1 sample we use data for 12 nearby rich clusters, the same
sample analyzed in detail in RdS98. This nearby sample consists of
clusters with $z < 0.055$ having at least 65 members, within 2.5 \Mpc
from the cluster center, morphologically classified as elliptical,
spiral or lenticular galaxies.  Clusters with obvious substructures
were discarded, since in those cases the velocity distribution could
result from a complex association of several small groups.

Because of the non-homogeneity of the spatial distribution in the
CNOC1 sample (i.e. galaxy members selected along elongated strips), we
decided to work with data inside the fiducial radius r$_{200}$.  This
radius | defined as the radius where the mean interior density is 200
times the critical density of the universe | is expected to contain
the bulk of the virialized cluster mass.  To derive r$_{200}$ from the
observational virial radius, $r_v$ (which is largely fixed by the
outer boundary of the sample), we assume that $M(r)\propto r$. This
gives

\begin{equation}
r_{200} = \frac{\sqrt{3} \sigma}{10 H_o (1 +
z) \times (1 + \Omega_o z )^{1/2}} 
\end{equation}

\noindent which is completely independent of the observational virial
radius (Carlberg, Yee \& Ellingson 1997).  Here $\sigma$ is the global
velocity dispersion of each cluster. A Hubble constant of H$_0$ = 100
\kms \Mpc$^{-1}$ and $\Omega_0 = 0.2$ are adopted.  An iterative
procedure was applied to select the members inside the r$_{200}$
region of each cluster. First an initial value of the velocity
dispersion was assumed, using values from Carlberg et al. (1996) for
the CNOC1 data and from RdS98 in the case of the nearby sample.  The
r$_{200}$ was estimated and galaxies inside this radius were selected,
calculating a new velocity dispersion.  The procedure was stopped if
the value for the velocity dispersion of the selected galaxies in two
subsequent iterations did not differ by more than 50 \kms. No more
than 3 iteration were necessary for each of the clusters.

The properties of both samples are presented in Table 1, where the
clusters are ordered from high to low redshifts. Column (2) presents
the number of galaxies inside r$_{200}$ and column (3) shows r$_{200}$
in \Mpc.  The mean redshifts and the velocity dispersions (columns 4
and 5), and all other means and dispersions appearing in this paper, 
were estimated using the bi-weighted estimators described in Beers et
al. (1990), using the ROSTAT\footnote{Version obtained from the ST-ECF
Astronomical Software Library ftp://ecf.hq.eso.org/pub/swlib} program,
which contains the versions of statistical routines tested by T.
Beers, K.  Flynn, and K. Gebhardt for robust estimation of simple
statistics. Also, all the errors bars are at the 68\% confidence
level, and they were obtained via a bootstrap resampling procedure
with 1000 iterations. Column (6) presents the cluster concentration
parameter as was defined by Butcher and Oemler (1978), i.e. the
logarithm of the ratio between the radius that contains 60\% and 20\%
of the cluster population, $C = log_{10}(r_{60}/r_{20})$.  Finally, in
column (7) clusters with obvious substructure are marked with a letter
{\it a} and those that do not cover the whole region inside r$_{200}$ are
marked with a {\it b}, all these clusters were not considered in the
final sample.
 
\subsection{Morphological classification}

The use of ground-based imaging for measuring morphology of
high-redshift galaxies is discussed in Schade et al. (1996, hereafter S96)  
where
details are given of the procedure, including convolution with the
point-spread function. Although it is difficult to perform Hubble type
morphological classification even at moderate redshifts, it is
possible to define roughly 3 galaxy classes based on the fractional
bulge luminosity (B/T) after fitting two-component models
(de Vaucouleurs $R^{1/4}$ bulge and exponential disk).  The data quality
here is lower than that used in S96 but the redshifts
are lower, on average on that a similar level of morphological
discrimination is feasible.  The correspondence between the ratio B/T
and morphological type was established by using the following recipe:
(i) disk-dominated profiles, i.e.  with $0.0 \leq$ B/T $\leq 0.4$,
were defined as late-type galaxies, (ii) profiles with $0.4 <$ B/T $<
0.7$ were defined as intermediate-type galaxies, and (iii)
bulge-dominated profiles, i.e. $0.7 \leq$ B/T $\leq 1.0$ were defined
as early-type galaxies.

The distribution of the rest frame color index (U-V)$_0$ of each
morphological class in the CNOC1 cluster galaxies are presented in
Figure 2.  The difference between the early- and late-type color
distributions shows that morphological cuts that we have defined
provides some level of real discrimination between galaxy types since
the profile fits are independent of color. As expected from local
samples of galaxies, objects classified as early-type galaxies from
the B/T indicator are redder than those classified as late-type
galaxies. In fact, the early-type galaxies have a mean color index
(U-V)$_0$ = 1.84 $\pm$ 0.01 with a bi-weighted dispersion of 0.17.
This distribution present a small color scatter, as usually found in
early-type galaxies at intermediate-redshifts.  However, the late-type
present two peaks: a blue distribution centered on (U-V)$_o$ = 1.06
$\pm$ 0.03 with a bi-weighted distribution of 0.21, and a red
distribution centered on 1.82 $\pm$ 0.01 (bi-weighted dispersion of
0.21).  Then, the population of these late-type galaxies could be
further separated into two groups: a red disk-dominated galaxies, and
a blue disk-dominated galaxies, which could represent the very
late-type (i.e. Sd or Irr).  This result shows that in the late-type
case, when the B/T ratio is considered, we are not separating the
galaxies by their star-formation activity.  Instead, we are separating
them by how well they are represented by either an exponential and/or
a de Vaucouleurs profile, and this is not related with their internal
kinematical state, because virialized systems could present both types
of profiles.  A comparison of the late-type color distribution with
Figure 6 of S96 indicates a significant difference in color between
disks in clusters and the fields. It is seems that the bluer peak of
the disk-dominated galaxies in clusters correspond to the field
population. In another side, the bulge-dominated galaxies in clusters
seems to be clean of the star forming galaxies, as Im, Sbc observed at
the field. But, the last result is accentuated by the fact that from
the final 716 galaxies morphologically classified, inside the
$r_{200}$ in the 9 selected clusters, 15 were excluded as peculiar
galaxies (with $B/T > 0.5$ and color index in the rest-frame $(U-V)_0
< 1.4$), as is suggested in S96. Other 39 were also discarded because
their low quality profile fitting.
 
\section{Morphological segregation}

We combine the two samples of clusters, in normalized co-ordinates to
construct two ensemble working samples, called ec-nearby and
ec-cnoc. Figure 3 presents the projected positions of the galaxies
inside the r$_{200}$ radius of the ec-nearby and ec-cnoc clusters. To
combine the clusters we adopt the same procedure of Carlberg, Yee \&
Ellingson (1997) and Ram\'{\i}rez \& de Souza (1998).  The brightest
cluster galaxies were used as the nominal centers of each cluster on
the sky for the CNOC1 clusters.  The galaxy velocities were normalized
to the velocity dispersion about the cluster mean, and the projected
radii were normalized to the empirically determined r$_{200}$ (see
section 3).  This procedure diminishes substructure and asphericity to
a level where the galaxies can be treated as if they were in a
spherical distribution, preserving the radial dependence of the
kinematical properties of the samples, and clearing off the effects
due to eventually existing local substructures.  Before any comparison
between both ensembles, it is important to clarify how the different
morphological populations will be referred. Due to the selection of
only elliptical, lenticular and spiral galaxies in the nearby sample,
hereafter the definition of early-type (late-type) galaxy will mean:
elliptical (spiral) galaxy in the nearby sample and bulge-dominated
(disk-dominated) galaxy at intermediate-redshift. Then, the early-type
galaxies not include the lenticular galaxies in the nearby sample, nor
the intermediate B/T class galaxies at intermediate-redshift.  Also,
we must to note that the late-type definition in the nearby clusters
excluding the irregular galaxies.

\subsection{Projected position distribution}

From the Figure 3 we readily noted that the spatial distributions of
the late-type galaxies are broader than those for the early-type
galaxies, in both samples. This morphological segregation had already
been detected in the nearby sample and seems to be also present in the
CNOC1 data.  A useful quantitative parameter to compare the spatial
distribution of each morphological class is the concentration C,
listed and defined in Table 1.  We found that late-type galaxies tend
to present a mean concentration lower than that for early-type
galaxies.  The typical values for the late-type galaxy populations
are: C$_{S-nearby} = 0.40$ and C$_{S-cnoc} = 0.31$. For the more
concentrated early-type galaxies the values are C$_{E-nearby}=0.59$
and C$_{E-cnoc}= 0.42$, and for the intermediate-type it is
C$_{S0-nearby} =0.50$ and C$_{S0-cnoc}= 0.41$.  Although it is clear
that the concentration parameter is higher for the early-type galaxies
in both samples the effect is less pronounced in the case of the
intermediate-redshift clusters.

Figure 4 shows the  morphology-radius relation that galaxies separated
by the B/T  criteria  follow.  This figure shows that the
bulge-dominated  galaxies are   centrally   concentrated   while the
disk-dominated population prefers the periphery. This simply shows the
morphology-density (Dressler 1980) or morphology-radius (Whitmore, 
Gilmore, \& Jones 1993) relation over a wide range in cluster-centric 
distance. The field galaxies included in the CNOC1 survey are also 
included in the plot.

\subsection{Kinematical segregation}

Figure 5 presents a summary of several cluster properties as a
function of $z/(1+z)$.  For low values of redshifts $z/(1+z)
\rightarrow z$, and it is proportional to the look-back time ($\tau$)
when $\Omega_0 \rightarrow 0$. In fact, $\tau = z/(1+z) H_o^{-1}$ Gyr
when $\Omega_0 = 0$.  We prefer this variable because the clusters
with redshifts lower than 0.1 lay more separated in a plot with this
scale than in a linear z scale, which is also labeled in the upper
axis.

Figure 5a shows the distribution of the radius r$_{200}$ in \Mpc.  The
dotted and dashed lines represent the r$_{200}$ of the Coma and Virgo
clusters, respectively.  None of the CNOC1 clusters is as massive as
Coma and three of them have significantly low masses.  Figures 5b
and 5c show the ratio between the velocity dispersion of late- and
early-type galaxies, $\sigma_S/\sigma_E$, and the ratio between
intermediate and early-type galaxies, $\sigma_{S0}/\sigma_E$,
respectively. As can be seen late-type galaxies at low and
intermediate-redshifts show similar values, while the intermediate
galaxies display values from $\sigma_{S0}/\sigma_E \sim 0.50$ to
$\sim$1.8 in both samples.  It has previously been suggested that
late-type galaxies are just now being captured while early-type
galaxies may have been in the cluster since the initial formation,
representing the relaxed and virialized population, respectively (see
references in section 1).  Then, a value of $\sqrt 2$ must be expected
(dot-and-dashed line in the Figure 5b and 5c), however in most
clusters the ratio is lower. Actually, the mean values are
$\sigma_S/\sigma_E = 1.14 \pm 0.07$ (bi-weighted dispersion of 0.18)
for the nearby sample and $\sigma_S/\sigma_E =1.17 \pm 0.06$
(bi-weighted dispersion of 0.18) for the CNOC1 data. The velocity
dispersion of the late-type galaxy population is approximately 15\%
higher than the velocity dispersion for the early-type galaxies.
These results show the existence of a kinematical segregation in both
samples | nearby and intermediate-redshift | where the population of
early-type galaxies always have a lower velocity dispersion than the
population of late-type galaxies. The similar values along the
redshift suggest that no-evolution for this type of segregation is
present.

Figure 5d shows the distribution of the concentration as a function of
redshifts, where the dotted line represents the Coma cluster
concentration, and the dashed line correspond to the Virgo cluster
concentration, and open diamonds show the concentrations of nine
Dressler et al.  (1997) (hereafter D97) clusters.  D97 measured the
concentration of 10 clusters at intermediate-redshifts, using galaxies
well inside the central region that cover $\sim$ 0.5-0.8 \Mpc
(i.e. inside the corresponding $r_{200}$ if they have $\sigma < 1200$
\kms). They re-evaluated the T-$\Sigma$ relation, using these
clusters, finding that the more concentrated clusters are the most
affected by the T-$\Sigma$ relation.  We observe that most of the
nearby clusters have similar concentration values | i.e. concentration
around 0.42-0.52 | and they are more concentrated than the
intermediate-redshift cluster. The CNOC1 clusters appear with
concentrations similar to the low concentration cluster of D97 sample.
Then, in agreement with the conclusions of D97 we must expect a lower
morphological segregation within the CNOC1 clusters.

\subsection{Lenticular galaxies evolution}

Finally, in the last two plots of Figure 5 we detected evolution
comparing the numbers of late- and intermediate-type to early-type
galaxies. In this case we are based in the fact that early-type
galaxies must correspond to the oldest population in the cluster |
i.e. if they were formed by merging, that must have happened at the
beginning of the cluster formation, or if they were formed in a
monolithic way they were formed all together not after $z = 2$ | and
we are also assuming that their numbers kept constant along the
cluster life-time.  In D97 the ratio of lenticular to elliptical
galaxies is cited as an indicator of evolution, and they suggest this
ratio decreases with redshift.  This is the same tendency found in our
samples, as Figure 5f shows.  Although, when we compare our data
(filled circles in Figure 5f), with those of D97, where a visual
morphological classification with HST images was used (open diamonds
in Figure 5f), we observe an underestimation of our number of
intermediate-type galaxies. This is not only an effect of the
difference between morphological classification methods, because the
nearby sample | where the classical visual morphology classification
was also used | follow the same tendency of the CNOC1
intermediate-redshifts clusters.  Nevertheless, in all the cases the
number of intermediate-type galaxies is significantly lower at higher
redshifts.  In fact, the mean value of $N_{S0}/N_E$ is $1.46 \pm 0.34$
for the nearby sample and $0.43 \pm 0.04$ for the CNOC1 sample.
Unfortunately, the physical meaning of this evolution is still in
discussion and the problem with the misidentification between
elliptical and lenticular galaxies is always present.  Interesting is
to note that when the late-type galaxies are compared with the
early-type galaxies the ratio $N_{S}/N_E$ (Figure 5e) is almost
constant, and they have similar values of that of D97 sample.

\section{Segregation detected in the average deviation of the 
velocity distribution}

The average deviation $|u|$ defined in section 2, is useful to connect
the line-of-sight velocity distribution of a population with the
family of orbits they represent. RdS98 analyzed the behavior of this
parameter for a sample of 18 nearby clusters, and concluded that
elliptical galaxies have more eccentric orbits than spiral galaxies.
In this section we investigate if the results found for the nearby
cluster sample also apply for the CNOC1 sample.

The velocity average deviation normalized by the velocity dispersion,
$|u|$, was calculated for each population class as follow:

\[ |u| = \frac{1}{N} \sum_{i=1}^N |v_i-v_{cl}|/\sigma_{cl} \]

\noindent where N is the number of galaxies of each morphological
class, $v_i$ is the line-of-sight velocity, $v_{cl}$ and $\sigma_{cl}$ 
are the mean velocity and velocity dispersion of the cluster, 
and they were calculated with the bi-weighted estimators (Beers et al 1990).
 Table 2 presents the average deviation and other kinematical properties of all
morphological classes in the nearby and CNOC1 clusters.  Column (1)
shows the name of the cluster, in the case of Virgo the first entry
(Virgo-all) corresponds to the whole cluster, considering as a part of the cluster
all the substructures detected by Binggeli, Popescu \& Tammann (1993), inside $r_{200}$, 
the second entry (Virgo-A) corresponds to the main cluster only, part A. 
The columns (2), (3) and (4) present
the following properties for the late-type galaxies: number of
galaxies inside the r$_{200}$ radius, velocity dispersion normalized
by the cluster velocity dispersion, and average deviation also
normalized by the velocity dispersion with their errors to the 68\%
confidence level.  Columns (5), (6) and (7) show the same properties
described above for intermediate-type galaxies and columns (8), (9)
and (10) for early-type galaxies.  When less than 6 galaxies were
available the dispersion and the average deviation were not
calculated. In Figure 6 we plot the distribution of the average
deviation for the late-, intermediate-, and early-type galaxies as
a function of the same variable $z/(1+z)$ used in Figure 5.  The
dotted lines show the expected value of $|u| = 0.77$, which represents
the highest value bellow which an eccentric orbit can be unambiguously
identified, as was described in section 2. For clusters below the dotted 
we should have that the radial velocity dispersion is larger than the
transversal velocity dispersion, $\sigma_R > \sigma_\perp$, under the
assumption of Gaussian velocity field adopted in the present study.

There is a possible bias toward higher values of average deviation due 
to substructures within clusters, because an artificial enhancement is
introduced in the projected velocity distribution.
This bias could be more important in the intermediate redshift sample,
where substructures are hardly detected.  
Cen (1997) using n-body simulations suggested that the presence of
substructures modifies the velocity distribution in a complex way, but
the  final  velocity   dispersion  is slightly    affected. He estimated
variations of the final velocity dispersion  of only 5\% and  9\% 
within 0.5 and  1.0 \Mpc, respectively.  Another study is presented 
in Bird et al. (1996),  using observational data, where they noted that
the existence of substructures is an important factor to determine the
dynamical  parameters, but the  effect  is reduced  when the data  is
restricted to a region inside the  virial radius, as we made.
 It is worthwhile to mention that variations
on the velocity dispersion due to substructures could increase or reduce
the velocity dispersions (Bird,  1994), a  fact which is   
not consistent with spirals always presenting an slightly larger   
dispersion than   ellipticals. Then, the effect produced by substructures 
seems to be weak enough to allow the segregation be detected. 
However, to test how the average deviation could be modified by 
substructures we applied the method to the Virgo cluster, 
taking into account the substructures detected by Binggeli, 
Popescu \& Tammann 1993, (BPT93). 
First we worked with the main body of Virgo, called part A
in BPT93, then the same method was applied including
the part B and the clouds W and W'. All samples were 
limited to the region inside the r$_{200}$ radius. 
The results are presented in the two Virgo's entry in the 
Table 2.  Both samples show 
very similar results and  differences between 
spiral, lenticular and elliptical populations are detected
on spite of the substructures.

\subsection{Late and early-type galaxies}

Figure 6 shows the segregation at low redshifts previously detected by
RdS98 (see their Figure 4), but now considering galaxies inside
r$_{200}$ radius instead of the limiting radius of 1.0 \Mpc and 2.5
\Mpc adopted in that analysis.  The segregation at low redshift shows
that elliptical galaxies have $|u|$ values corresponding to highly
eccentric orbits, while the spiral galaxies have orbits that are
nearly isotropic, circular or slightly radial.  Although in general
the population of early-type galaxies in the CNOC1 clusters follow the
same tendency of those of the nearby clusters, having also lower
average deviation values, the clear segregation detected for clusters
at $z < 0.1$ is not so obvious at higher redshifts.  This result could
indicate a real evolution of the orbit shapes inside the $r_{200}$
radius, where the early-type population only recently got more
eccentric in their orbits, and the population of late-type galaxies
kept in circular or isotropic orbits.  However, the distribution of
average deviations of early-type galaxies in the CNOC1 sample have
cases as ms1008-12 with orbits as eccentric as the nearby elliptical galaxies.
We must note that the intermediate-type galaxies in the nearby sample
follow the same behavior as the early-type population.  However, at
intermediate redshift the small number of intermediate-type galaxies,
producing a behavior not well defined in their kinematical parameters.

To further quantify the degree of significance of the differences
between the late-type and early-type galaxies along the redshift,   we have
applied   statistical  tests.  The  main goal  is to verify  if the
observed distributions  of the average  deviations in Figure  6,  could
come  from the same  parent distribution. The same  two-sampling tests
were  applied  to  all   clusters. The  tests applied  were developed
in IRAF/STSDAS ({\em  twosampt} and {\em kolmov}). The results are
presented in Table 3, where columns show: (1)  the two population
being  compared, (2) the  sample and the morphological criteria, (3)
the number of clusters used,  and (4) the mean, the error and the 
dispersion of the average  deviation distribution of each sample. 
The  four last columns present the probability, expressed in percentage,  
of the two compared
population belong to  the  same distribution,  using  the:  (5)  K-S
Test,  (6)  Gehan's Generalized Wilcoxon  Test, (7) the Logrank  Test
and the  (8) Peto \& Prentice  Generalized Wilcoxon  Test.   The
statistical  tests show  a significant difference  between the average
deviation distribution of the  local spiral and  elliptical galaxies
 at the 99\% of confidence level. In the case of bulge- and disk-dominated  
galaxies at intermediate-redshifts the distributions of the average 
deviation are different at the 67-89\% of confidence level, 
depending of the statistical test. Then, the difference in the
velocity anisotropies seems to exist but is weaker than those for 
the nearby galaxies.

\subsection{Red and blue galaxies}

Carlberg et al. (1997) compared the properties of the CNOC1 galaxies
separating them by colors. They applied the Jeans equation of stellar
hydrodynamical equilibrium to the red and blue subsamples, and they
found that both populations give separately statistically identical
cluster mass profiles.  This is strong evidence that the CNOC1
clusters are effectively systems in equilibrium.  Motivated by this
work we have also repeated the analysis separating the galaxies by
colors.  Unfortunately we do not have access to homogeneous data on
colors for the nearby clusters. A comparison between nearby and
intermediate-redshift clusters requires complete data color for both
samples with the same set of filters. Therefore the comparison between
red and blue galaxies is applied to the CNOC1 data only.

The criteria to separate the red and blue galaxies was the following:
(i) galaxies with (U-V)$_0 \leq 1.4$ were defined as blue galaxies,
associated to the late-type galaxies, (ii) galaxies with (U-V)$_0 >
1.4$ were defined as red galaxies, associated to the early-type
population.  Applying our orbital analysis to these two subsamples we
find that the average deviation of both populations, in all clusters,
have very different distributions (different at the 99\% confidence
level).  The last line in Table 3 shows this result and Table 4 shows
in detail the properties of these populations. In this case we are
working with only 7 of the 9 clusters because ms1455+22 and ms0440+02
have less than 6 blue galaxies, and no-statistical results are
possible with this number of objects. The mean average deviation of
the blue population is $|u| = 1.02 \pm 0.09$ (dispersion of 0.21) and
for the red population is $|u| = 0.70 \pm 0.02$ (dispersion of
0.06). Therefore, similarly as occurred with nearby ellipticals, the
red galaxies observed at intermediate-redshift also have more
eccentric orbits than the blue galaxies.  In addition, we observe that
the mean velocity dispersion of the red and blue galaxy populations |
calculated with the velocity dispersion of each population normalized
to the velocity dispersions of the respectively cluster | are 0.96
$\pm$ 0.04 and 1.29 $\pm$ 0.09, respectively. That is, the velocity
dispersion of blue cluster galaxies is about 30\% higher than for red
galaxies.  This quantity is larger than the difference between
elliptical and spiral at low redshift and this one of disk- and
bulge-dominated galaxies at intermediate-redshift. 

Before any consideration about the notable difference between red and
blue galaxy population inside $r_{200}$, we must mention two important
characteristic of these subsamples.  First, the difference detected
here is not in contradiction with the conclusion of Carlberg et
al.(1997) about red and blue galaxies in the CNOC1 survey being
systems in equilibrium with the gravitational potential of their own
clusters. As was explained in RdS98, the global parameters as velocity
dispersions are independent of the existence of any difference in the
velocity anisotropy. And second, unfortunately up to the limit of
$M_r^k=-18.5$ mag, about 85\% of the cluster galaxies inside r$_{200}$
fall into the red subsample.  This effect produce a limiting factor in
the precise numerical agreement because there are relatively few blue
galaxies in clusters, meaning that the mean velocity dispersion and
average deviation are less accurately measured than those for red
galaxies.  On spite of the few galaxies, the behavior of the blue
population in all the clusters is always in the same direction (see
Table 4).  If the blue galaxies are galaxies rich in gas they could
represent the star formation activity of the cluster.  This activity
could be due to an internal perturbation (e.g. early stage of the gas
consumption), external perturbations (e.g. tidal effect by near
companions or recent merging), or due a mix of these two (e.g.
internal response to the galactic harassment).  Define how all of
these perturbations will affect the observed stellar distribution is a
hard task. However, the orbital properties can produce measurable 
properties compatible with the star formation activity.  The objects
with very eccentric orbits reach the densest regions of the cluster
and are more prompt to suffer perturbations from other
galaxies. Although the high relative velocities of the galaxies in the
cluster center makes merging unprovable, the accumulated effect of
these perturbations could remove the gas content of the galaxies,
dimming the star formation rates (e.g. Fujita et al. 1999).  In this
cases, the blue galaxies in clusters tends to appear redder after some
crossing times due to the velocity anisotropy.  This could be the
case of the A2390 cluster studied by Abraham et al. (1996).  In fact,
they found that the blue galaxies in this cluster are being altered by
the cluster environment, such that some of the their members are
likely leaving the blue population to join the red population.  Then,
if the other clusters in the CNOC1 survey have velocity anisotropy, as
A2390 cluster, our finding that red galaxies have more eccentric
orbits than blue galaxies is consistent with a modification of the
star formation activity by an orbital-segregation.

\section{Conclusion}

We presented a kinematical analysis of clusters at
intermediate-redshifts. They were analyzed looking for the existence
of an orbital-segregation, in the sense that galaxies with different
morphological types have different orbit families, as was previously
found for ellipticals and spirals in nearby clusters.  The velocity
anisotropies of each morphological population was estimated by the use
of the average deviation of the line-of-sight velocity, which is
associate to the orbital shapes, assuming a Gaussian distribution
having different dispersions along the radial and the transversal
directions.

The differences in the average deviation of velocity distributions of
elliptical, lenticular and spiral galaxies detected inside the 2.5
\Mpc in nearby rich clusters (RdS98) is stronger inside the more
interesting in physical meaning $r_{200}$ radius.  These differences
correspond to different anisotropies and it corresponds to an stronger
orbital-segregation.  In this case the early-type galaxies, inside the
$r_{200}$ radius, have orbits more eccentrics than the late-type
galaxies, and the intermediate-type galaxies share an intermediate
orbital family.

When the procedure applied to the nearby clusters is applied to a
sample of intermediate-redshifts cluster, as the CNOC1 survey, an
orbital-segregation is again detected.  However, this time the
differences is between the bulge-dominated galaxies and the
disk-dominated galaxies, and the orbital-segregation is less strong
than in the nearby sample. Moreover, when the orbits of the red and
blue galaxies of the intermediate-redshift sample are compared the
strongest orbital segregation is found. This result suggests  the orbital
segregation seems to modify more efficiently the star formation
activity than the internal shape of the galaxies.

Then, our results impose a further restriction on the plausible models of
cluster of galaxies formation.  Along the covered redshifts the models
 must reproduce the observed velocity field
anisotropy of each morphological type.  In this case the early-type
galaxies are evolving very fast from z=0.4 to 0, and late-type
galaxies remain with their orbits without obvious evolution.  However
we cannot discard the possibility that this could be an effect of
comparing different population when we associate the elliptical
galaxies in nearby clusters to the bulge-dominated galaxies at
intermediate-redshifts.  Unfortunately, in the case of
intermediate-type galaxies, their behavior at higher redshifts is not
so clear as in the nearby cluster, and we have | besides the always
present problem of misidentification | the problem of the dramatic
decrease in number.

Finally, some interesting question are opened:
Does the orbital-segregation represent the tendency of early-type 
galaxies to preserve the initial conditions of an anisotropic
collapse?  In effect, this type of collapse studied with high
resolution n-body simulations and semi-analytical models forms
clusters with an anisotropic velocity distribution of their members.
These clusters are different from those obtained in a classical
spherical collapse, without taking into account the anisotropy in
large-scale initial conditions. Cosmological n-body simulations have
shown that, in the most scenarios, the matter in filaments collapses
into dark matter halos, and these flow along filaments towards the
potential minima, where they form galaxy clusters. The collapse is
therefore a highly anisotropic process (West, Villumsen \& Dekel 1991,
Tormen 1997, Tormen, Diaferio \& Syer 1998; Ghigna et al. 1998;
Dubinski 1998; Splinter et al. 1997). For example, Tormen (1997)
studying cluster formation with n-body simulations found that more
massive satellites move along slightly more eccentric orbits, which
penetrate deeper in the cluster.  Also, he found that the shape and
orientation of the final cluster and its velocity ellipsoid are
strongly correlated with each other and with the infall pattern of
merging satellites, suggesting that cluster alignments is related to
the anisotropy in the velocity. In addition, the properties of dark
matter halos within rich clusters studied by Ghigna et al. (1998) 
present an orbital distribution close to isotropic,
however the circular orbits are rare and the radial orbits are
common. Although, the interesting results cited above are more
representative of the dark-matter distribution, few has been done to
compare with real clusters. Only some observations of distribution
of structure in X-ray clusters (West, Jones \& Forman 1995) seem
indicate that anisotropic collapse is common in also in real clusters.
Nevertheless, why only elliptical galaxies are able to preserve the
initial conditions? And why the youngest sample shows
very eccentric orbits only for red galaxies?  Could this be the result
of a morphological transformation within an anisotropic collapse?  In
this case, how the strength of this transformation will depend of the
galaxy orbit?.  That question, must be studied and clusters are higher
redshift must be studied.

\acknowledgments

We thank to the members of the CNOC team who providing us with 
the data, in special we thank the comments from Ray G. Carlberg. 
Also AR thanks Claudia Mendes de Oliveira who helped her with
comment that led to substantial improvements in the presentation.
Also we gratefully acknowledge financial support  
from FAPESP (AR Post-Doc fellowship grant No 1998/014345-9)
and  FAPESP (RES grant No 1995/7008-76). And RES acknowledge 
support for this work provided by the National Science
Foundation through a Gemini Fellowship from the Association of 
Universities for Research in Astronomy, Inc., under NSF cooperative 
agreement  AST-8947990.


\clearpage

\begin{deluxetable}{lrccrrcl} 
 
\small
\tablenum{1}
\tablewidth{0pc}
\tablecaption{Cluster kinematic parameters from galaxies inside $r_{200}$}
\tablehead{
\colhead{Name}          &
\colhead{N}             &
\colhead{$r_{200}$}     &
\colhead{~~~~~~~~~z~~~~~~~~~}             &
\multicolumn{2}{c}{$\sigma _{r_{200}}$}& 
\colhead{~~~~~~C~~~~~~}     &
\colhead{Note} \nl
\colhead{}    &
\colhead{}    &
\colhead{(\Mpc)} &
\colhead{}&  
\multicolumn{2}{c}{(\kms)}& 
\colhead{}    & 
\colhead{}    \nl
\colhead{}    &
\colhead{}    &
\colhead{} &
\colhead{}&  
\multicolumn{2}{c}{-------------------}& 
\colhead{}& 
\colhead{} \nl
\colhead{}    &
\colhead{}    &
\colhead{} &
\colhead{}&  
\multicolumn{2}{c}{Value ~~~ Error}& 
\colhead{}& 
\colhead{} \nl
\colhead{(1)}&
\colhead{(2)}&
\colhead{(3)}&
\colhead{(4)}&  
\multicolumn{2}{c}{(5)}& 
\colhead{(6)}& 
\colhead{(7)} \nl
}
\startdata
{\bf CNOC1}   &     &     &        &        &       &        &        \nl
ms0016+16     &  73 & 1.5 & 0.5473 & 1449   &  124  &   0.44 &{\it a} \nl
ms0451-03     &  59 & 1.6 & 0.5384 & 1461   &  116  &   0.37 &{\it a} \nl
ms1621+26     &  43 & 1.0 & 0.4259 &  862   &   86  &   0.39 &        \nl
ms0302+16     &  25 & 0.9 & 0.4243 &  769   &  112  &   0.49 &{\it b} \nl
ms1512+36     &  27 & 1.0 & 0.3716 &  857   &  124  &   0.38 &{\it b} \nl
ms1358+62     & 128 & 1.3 & 0.3274 & 1046   &   61  &   0.43 &        \nl
ms1224+20     &  27 & 1.0 & 0.3256 &  824   &   98  &   0.35 &{\it b} \nl
ms1008-12     &  72 & 1.3 & 0.3070 & 1043   &  107  &   0.34 &        \nl
ms1006+12     &  30 & 1.2 & 0.2604 &  907   &   86  &   0.40 &{\it b} \nl
ms1455+22     &  57 & 1.5 & 0.2566 & 1095   &  119  &   0.25 &        \nl
ms1231+15     &  65 & 0.9 & 0.2348 &  686   &   59  &   0.46 &        \nl
a2390         & 128 & 1.7 & 0.2283 & 1200   &   69  &   0.33 &        \nl
ms0451+02     & 111 & 1.6 & 0.2006 & 1103   &   81  &   0.34 &        \nl
ms0440+02     &  33 & 1.0 & 0.1966 &  698   &   86  &   0.29 &        \nl
ms0839+29     &  79 & 1.3 & 0.1931 &  933   &   85  &   0.46 &        \nl
ms0906+11     &  92 & 2.8 & 0.1712 & 1918   &  104  &   0.33 &{\it a} \nl
              &     &     &        &        &       &        &        \nl
{\bf Nearby sample}&&     &        &        &       &        &        \nl
a0754         &  57 & 1.3 & 0.0542 &  767   &   82  &   0.49 &        \nl
a1644         &  81 & 1.5 & 0.0472 &  905   &   86  &   0.37 &        \nl
a3376         &  61 & 1.3 & 0.0464 &  756   &   74  &   0.50 &        \nl
a1631         &  46 & 1.1 & 0.0464 &  654   &   72  &   0.45 &        \nl
a0119         &  75 & 1.4 & 0.0445 &  840   &   94  &   0.48 &        \nl
a0496         &  74 & 1.0 & 0.0327 &  615   &   70  &   0.65 &        \nl
a1656         & 453 & 1.7 & 0.0233 & 1033   &   40  &   0.51 &        \nl
a0194         &  51 & 0.8 & 0.0178 &  463   &   54  &   0.27 &        \nl
a805s         &  37 & 0.9 & 0.0144 &  511   &   58  &   0.48 &        \nl
a1060         &  92 & 1.1 & 0.0122 &  647   &   45  &   0.65 &        \nl
Fornax        &  50 & 0.6 & 0.0048 &  347   &   36  &   0.52 &        \nl
Virgo         & 322 & 1.4 & 0.0045 &  802   &   26  &   0.43 &        \nl
\enddata

\tablenotetext{{\it a}}{cluster with substructure in the velocity
field and/or in the projected position distribution.}
\tablenotetext{{\it b}}{cluster with few velocities measured inside
r$_{200}$.}

\end{deluxetable} 

\clearpage

\begin{deluxetable}{lrcc@{~~~~~~}rcc@{~~~~~~}rcc} 

\small

\tablewidth{0cm}
\tablenum{2}
\tablecaption{Cluster kinematic parameters by populations}
\tablehead{
\multicolumn{1}{c}{}                          &
\multicolumn{3}{c}{{\bf Disk-dominated or S}~~~~~~~~} &
\multicolumn{3}{c}{{\bf Intermediate or S0}~~~~~~~~~~~}  &
\multicolumn{3}{c}{{\bf Bulge-dominated or E}} \nl
\multicolumn{1}{c}{Name}                            &
\multicolumn{1}{c}{N}                   &
\multicolumn{1}{c}{$\sigma _{S}/\sigma _{cl}$ }     & 
\multicolumn{1}{c}{$|u|$~~~~~~}                   &
\multicolumn{1}{c}{N}                   &
\multicolumn{1}{c}{$\sigma _{S0}/\sigma _{cl}$ }    & 
\multicolumn{1}{c}{$|u|$~~~~~~}                   &
\multicolumn{1}{c}{N}                   &
\multicolumn{1}{c}{$\sigma _{E}/\sigma _{cl}$ }     & 
\multicolumn{1}{c}{$|u|$}                   \nl
\multicolumn{1}{c}{(1)} &
\multicolumn{1}{c}{(2)} &
\multicolumn{1}{c}{(3)} &
\multicolumn{1}{c}{(4)~~~~~~} &
\multicolumn{1}{c}{(5)} &
\multicolumn{1}{c}{(6)} &
\multicolumn{1}{c}{(7)~~~~~~} &
\multicolumn{1}{c}{(8)} &
\multicolumn{1}{c}{(9)} &
\multicolumn{1}{c}{(10)}  \nl
} 
\startdata
          &     &        &                  &     &        &                  
&    &        &                   \nl
CNOC1     &     &        &                  &     &        &                  
&    &        &                   \nl
ms1621+26 &  21 &   1.07 &  0.70 $\pm$ 0.14 &   4 &     -  &    -             
& 17 &   0.97 &  0.80 $\pm$ 0.10  \nl
ms1358+62 &  41 &   1.06 &  0.85 $\pm$ 0.10 &  19 &   0.91 &  0.77 $\pm$ 0.11 
& 61 &   1.03 &  0.75 $\pm$ 0.08  \nl
ms1008-12 &  29 &   0.98 &  0.69 $\pm$ 0.11 &  12 &   1.30 &  0.96 $\pm$ 0.23 
& 25 &   0.70 &  0.46 $\pm$ 0.09  \nl
ms1455+22 &  22 &   1.14 &  0.80 $\pm$ 0.16 &  10 &   0.97 &  0.68 $\pm$ 0.21 
& 25 &   0.88 &  0.59 $\pm$ 0.10  \nl
ms1231+15 &  25 &   0.96 &  0.76 $\pm$ 0.11 &  12 &   1.31 &  0.96 $\pm$ 0.22 
& 28 &   0.90 &  0.55 $\pm$ 0.11  \nl
a2390     &  49 &   1.14 &  0.89 $\pm$ 0.09 &  22 &   0.77 &  0.56 $\pm$ 0.08 
& 47 &   0.86 &  0.65 $\pm$ 0.08  \nl
ms0451+02 &  45 &   1.04 &  0.78 $\pm$ 0.09 &  17 &   0.93 &  0.56 $\pm$ 0.16 
& 40 &   0.99 &  0.72 $\pm$ 0.08  \nl
ms0440+02 &  12 &   1.04 &  0.73 $\pm$ 0.20 &   9 &   0.51 &  0.15 $\pm$ 0.06 
& 10 &   1.14 &  0.89 $\pm$ 0.23  \nl
ms0839+29 &  38 &   0.93 &  0.57 $\pm$ 0.10 &  13 &   1.41 &  1.12 $\pm$ 0.22 
& 24 &   0.82 &  0.65 $\pm$ 0.10  \nl
          &     &        &                  &     &        &                  
&    &        &                      \nl
Nearby    &     &        &                  &     &        &                  
&    &        &                      \nl
a0754     &  18 &   1.05 &  0.67 $\pm$ 0.13 &  18 &   0.91 &  0.70 $\pm$ 0.12 
& 21 &   1.05 &  0.54 $\pm$ 0.15  \nl
a1644     &  19 &   1.08 &  0.72 $\pm$ 0.17 &  45 &   0.95 &  0.61 $\pm$ 0.07 
& 17 &   1.09 &  0.65 $\pm$ 0.15  \nl
a1631     &   6 &   1.03 &  0.82 $\pm$ 0.25 &  33 &   1.04 &  0.74 $\pm$ 0.10 
&  7 &   0.76 &  0.64 $\pm$ 0.24  \nl
a3376     &  19 &   1.00 &  0.75 $\pm$ 0.13 &  31 &   1.12 &  0.75 $\pm$ 0.12 
& 11 &   0.50 &  0.21 $\pm$ 0.07  \nl
a0119     &  15 &   1.14 &  0.94 $\pm$ 0.18 &  36 &   0.87 &  0.53 $\pm$ 0.09 
& 24 &   1.02 &  0.56 $\pm$ 0.16  \nl
a0496     &  16 &   1.05 &  0.72 $\pm$ 0.17 &  31 &   1.16 &  0.78 $\pm$ 0.14 
& 27 &   0.72 &  0.43 $\pm$ 0.09  \nl
a1656     & 188 &   1.05 &  0.73 $\pm$ 0.04 & 110 &   1.08 &  0.72 $\pm$ 0.05 
&155 &   0.88 &  0.59 $\pm$ 0.04  \nl
a0194     &  16 &   1.12 &  0.75 $\pm$ 0.15 &  19 &   0.73 &  0.53 $\pm$ 0.12 
& 16 &   1.06 &  0.73 $\pm$ 0.16  \nl
a805s     &  14 &   1.18 &  0.88 $\pm$ 0.20 &  15 &   0.85 &  0.69 $\pm$ 0.11 
&  8 &   1.01 &  0.58 $\pm$ 0.26  \nl
a1060     &  30 &   1.00 &  0.79 $\pm$ 0.11 &  49 &   0.99 &  0.76 $\pm$ 0.08 
& 13 &   0.92 &  0.47 $\pm$ 0.15  \nl
Fornax    &  13 &   1.16 &  0.93 $\pm$ 0.17 &  19 &   1.01 &  0.72 $\pm$ 0.13 
& 18 &   0.88 &  0.38 $\pm$ 0.16  \nl
Virgo-all & 196 &   1.06 &  0.85 $\pm$ 0.04 &  74 &   0.86 &  0.63 $\pm$ 0.05 
& 52 &   0.90 &  0.64 $\pm$ 0.07  \nl
Virgo-A   & 109 &   1.12 &  0.88 $\pm$ 0.05 &  55 &   0.76 &  0.56 $\pm$ 0.08
& 43 &   0.89 &  0.63 $\pm$ 0.06  \nl
          &     &        &                  &     &        &                  
&    &        &                      \nl
\enddata

\end{deluxetable}

\clearpage

\begin{deluxetable}{clrrrrrr}

\small
\tablenum{3}
\tablewidth{0pc}
\tablecaption{Two-sampling   test  between   populations}  
\tablehead{
\colhead{$Pop_1 \times Pop_2$} & 
\colhead{sample}            & 
\colhead{N$_{cls}$}         &
\colhead{$|u|_{Pop_1} \;\; \times \;\;  |u|_{Pop_2} $} &
\colhead{KS}                &
\colhead{GGW}               &  
\colhead{L}                 &   
\colhead{PPGW}              \nl
\colhead{}                  &
\colhead{}                  &  
\colhead{}                  & 
\colhead{}                  & 
\colhead{\%}                & 
\colhead{\%}                &
\colhead{\%}                &  
\colhead{\%}                \nl  
\colhead{(1)}               &  
\colhead{(2)}               &
\colhead{(3)}               &  
\colhead{(4)}               &  
\colhead{(5)}               &  
\colhead{(6)}               &
\colhead{(7)}               &  
\colhead{(8)}               \nl  
}  
\startdata  
{\bf  S  x  E}  &CNOC1$^{a}$ &  9 & 0.76 $\pm$ 0.03 (0.09) $\times$ 0.67 $\pm$ 0.05 (0.14) & 33   & 14   & 11   & 15   \nl 
                &Nearby$^{b}$& 12 & 0.79 $\pm$ 0.03 (0.10) $\times$ 0.56 $\pm$ 0.04 (0.14) &  0.01&  0.01&  0.00&  0.00\nl
                &            &    &                                                 &      &      &      &      \nl
{\bf S  x S0}   &CNOC1$^{a}$ &  9 & 0.76 $\pm$ 0.03 (0.09) $\times$ 0.74 $\pm$ 0.12 (0.31) & 63   & 77   & 23   & 76   \nl
                &Nearby$^{b}$& 12 & 0.79 $\pm$ 0.03 (0.10) $\times$ 0.70 $\pm$ 0.03 (0.09) & 10   &  1   &  1   &  1   \nl
                &            &    &                                                 &      &      &      &      \nl
{\bf  E x S0}   &CNOC1$^{a}$ &  9 & 0.67 $\pm$ 0.05 (0.14) $\times$ 0.74 $\pm$ 0.12 (0.31) & 63   & 50   & 96   & 50   \nl
                &Nearby$^{b}$& 12 & 0.56 $\pm$ 0.04 (0.14) $\times$ 0.70 $\pm$ 0.03 (0.09) &  3   &  1   &  1   &  1   \nl
                &            &    &                                                 &      &      &      &      \nl
{\bf Blue x Red}&CNOC1$^{c}$ &  7 & 1.02  $\pm$ 0.09(0.21) $\times$ 0.70 $\pm$ 0.02 (0.06) &  1   &  0.1 &  0.01& 0.04 \nl
                &            &    &                                                 &      &      &      &      \nl 
\enddata

\tablenotetext{{\it  a}}{S = (B/T$\leq$0.4), S0 = (0.4$<$B/T$<$0.7),  and 
E = (B/T$\geq$0.7)}

\tablenotetext{{\it  b}}{Hubble   type  morphology}

\tablenotetext{{\it  c}}{Blue = (U-V)$_0 \leq $1.4 and  Red = (U-V)$_0 > $1.4}

\end{deluxetable}

\clearpage

\begin{deluxetable}{lrccrcc} 
\small
\tablewidth{16cm}
\tablenum{4}
\tablecaption{Cluster kinematic parameters by populations} 
\tablehead{
\multicolumn{1}{c}{}  &   
\multicolumn{3}{c}{{\bf  Blue-(U-V)$_0  \leq 1.4$}}   &   
\multicolumn{3}{c}{{\bf   Red-(U-V)$_0   >   1.4$}}   \nl
\multicolumn{1}{l}{Name}    &        
\multicolumn{1}{l}{N}       &
\multicolumn{1}{c}{$\sigma_{blue}/\sigma_{cl}$}     &
\multicolumn{1}{c}{$|u|$}   &       
\multicolumn{1}{l}{N}       &
\multicolumn{1}{c}{$\sigma_{red}/\sigma_{cl}$}      &
\multicolumn{1}{c}{$|u|$}   \nl   
\colhead{(1)}  &  
\colhead{(2)}  &
\colhead{(3)}  &  
\colhead{(4)}  &  
\colhead{(5)}  &  
\colhead{(6)}  &
\colhead{(7)} \nl} 
\startdata 
ms1621+26 &  7 & 0.94 & 0.87 $\pm$ 0.18 &  36 & 1.05 & 0.78 $\pm$ 0.10 \nl 
ms1358+62 & 13 & 1.42 & 1.20 $\pm$ 0.17 & 115 & 0.95 & 0.70 $\pm$ 0.05 \nl 
ms1008-12 & 12 & 1.55 & 1.24 $\pm$ 0.27 &  60 & 0.86 & 0.56 $\pm$ 0.07 \nl 
ms1455+22 &  5 & -    & -               &  52 & 0.94 & 0.63 $\pm$ 0.08 \nl 
ms1231+15 &  6 & 0.47 & 0.76 $\pm$ 0.28 &  59 & 1.00 & 0.73 $\pm$ 0.08 \nl 
a2390     & 21 & 1.16 & 0.94 $\pm$ 0.13 & 107 & 0.97 & 0.72 $\pm$ 0.05 \nl 
ms0451+02 & 25 & 1.20 & 0.89 $\pm$ 0.13 &  86 & 0.95 & 0.68 $\pm$ 0.06 \nl  
ms0440+02 &  0 &    - &               - &  33 & 1.00 & 0.73 $\pm$ 0.11 \nl 
ms0839+29 &  7 & 1.50 & 1.25 $\pm$ 0.41 &  72 & 0.93 & 0.65 $\pm$ 0.07 \nl 
\enddata

\end{deluxetable}

\clearpage

\begin{figure} 
\centerline{\epsfxsize= 15cm \epsfbox{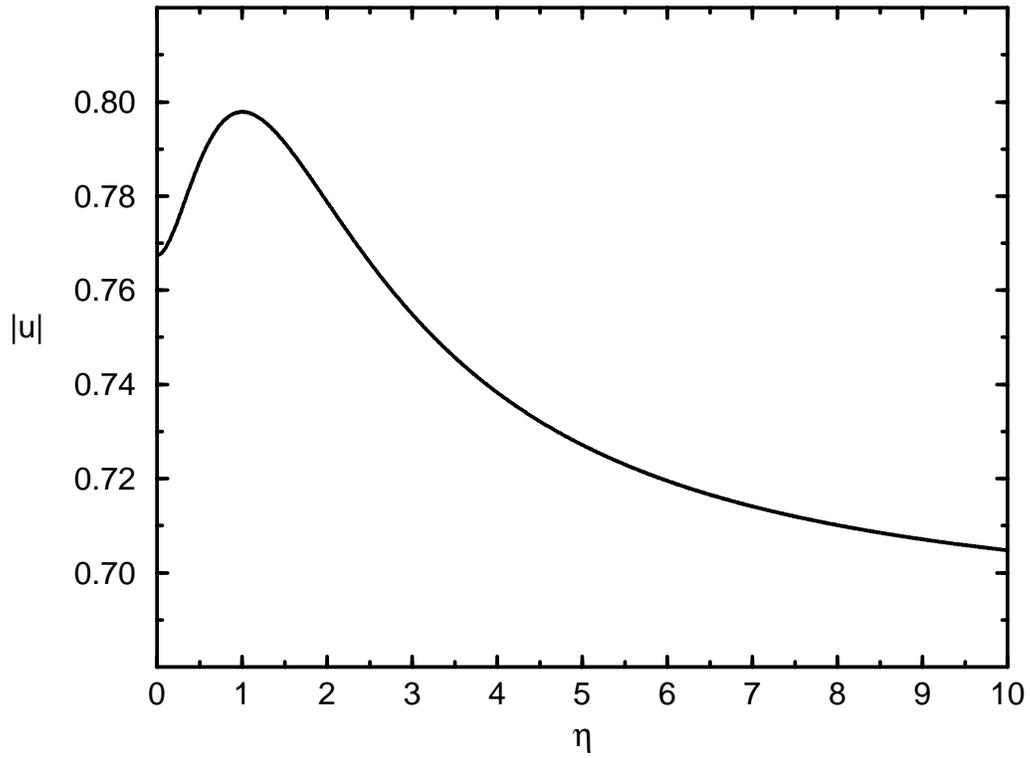}}
\caption{
Average deviation as a function of the anisotropic parameter
$\eta$. When the system have isotropic orbits ($\eta = 1$) the $|u|$
tends to the value 0.80. Systems with purely radial ($\eta >> 1$) or
circular orbits ($\eta \rightarrow 0$) have $|u|$ converging to 0.69
and 0.77, respectively.\label{f1} }
\end{figure}

\begin{figure} 
\vspace{1cm} 
\centerline{\epsfxsize= 15cm \epsfbox{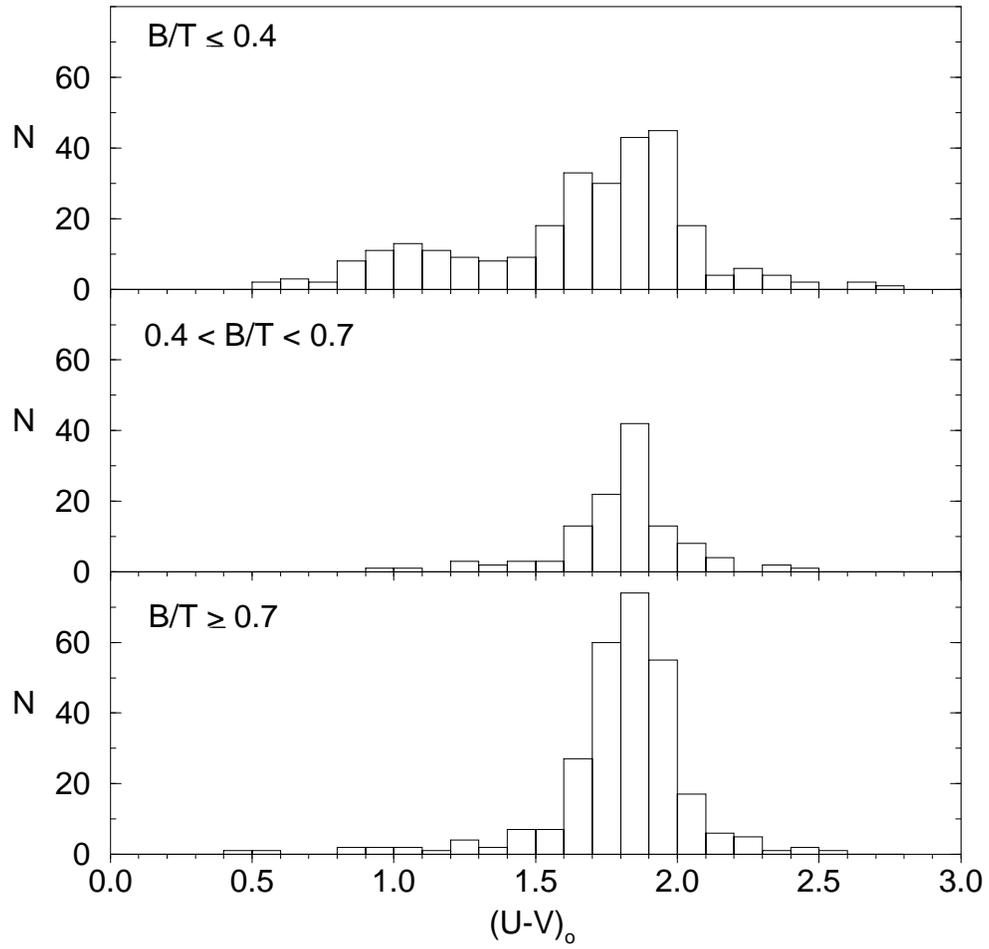}}
\caption{
Color distribution (U-V)$_0$ of the CNOC1 data when separated
by the B/T criteria. \label{f2}}
\end{figure}

\begin{figure}
\vspace{4cm}
\centerline{\epsfxsize= 8cm \epsfbox{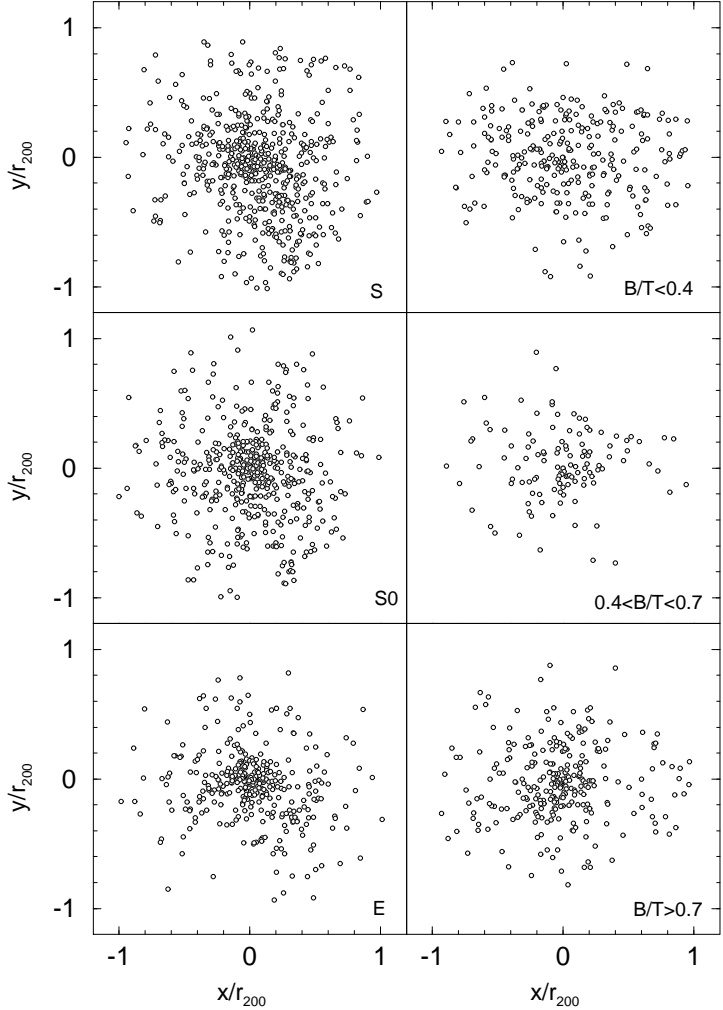}}
\vspace{5cm}
\caption{ 
Spatial distribution of galaxies in the ensemble clusters.
The projected radii of each galaxy is normalized to the $r_{200}$ of
the cluster it belongs to. Left, ec-nearby: projected positions of
galaxies in 12 rich nearby clusters. The positions are presented in
$r_{200}$ units, and they are separated as elliptical, lenticular and
spiral galaxies following the classical Hubble sequence morphological
classification.  Right, ec-cnoc: projected positions of galaxies in 9
intermediate-redshifts clusters, the galaxies were separated following
the B/T criteria (see section 3.1). \label{f3}}
\end{figure}

\vspace{8cm}

\begin{figure}      
\centerline{\epsfxsize= 15cm \epsfbox{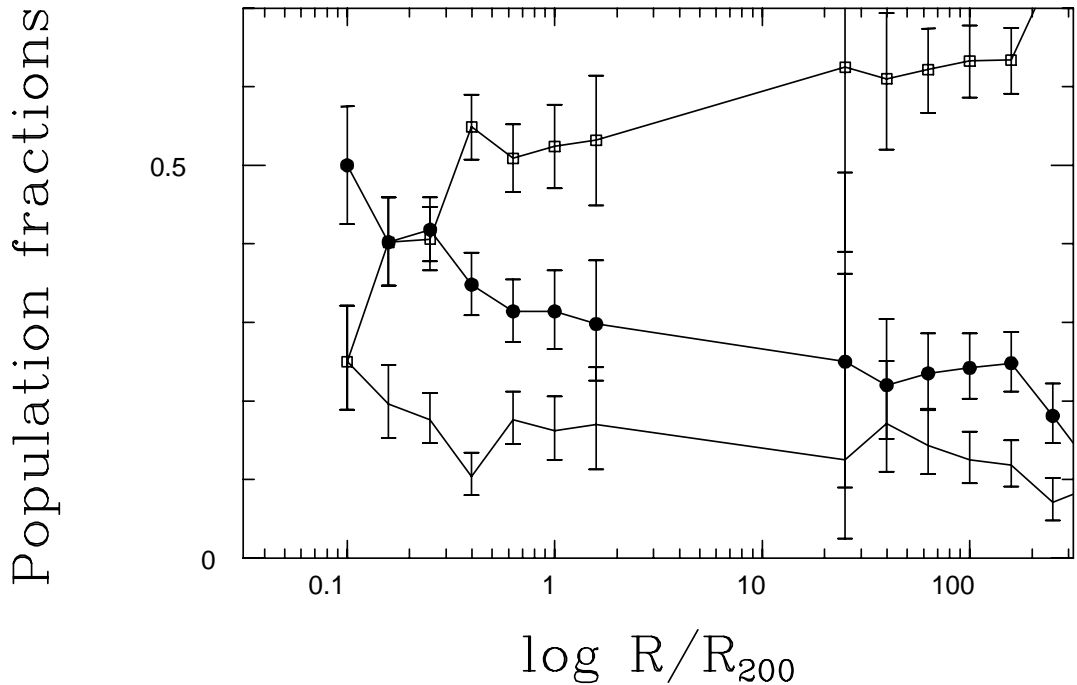}}
\caption{
Morphology-radius  relation of CNOC1  data.  Galaxies were
separated as bulge-dominant (B/T $>$ 0.7: fill circles), disk-dominant
(B/T $<$  0.4; open squares), and  intermediate galaxies  (0.4 $>$ B/T
$>$ 0.7; dots). The galaxies at $R/R_{200} > 2$ are field galaxies. \label{f4}}
\end{figure}

\begin{figure} 
\vspace{2cm}        
\centerline{\epsfxsize= 15cm \epsfbox{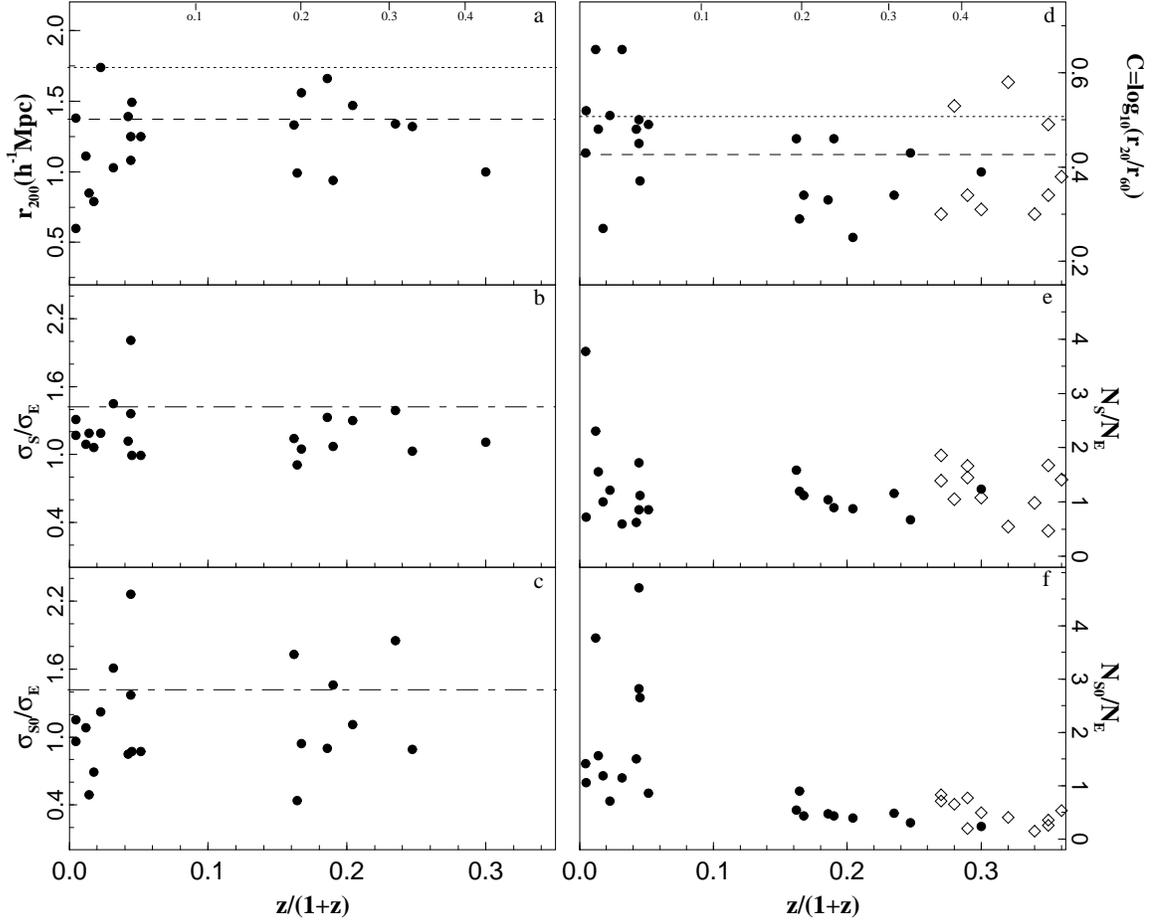}}
\caption{ 
Distribution along the variable $z/(1+z)$ of: (a) the radius $r_{200}$ 
in \Mpc, (b) the ratio between velocity dispersion of late- and early-type 
galaxies, $\sigma_S/\sigma_E$, (c) the ratio between velocity dispersion of 
intermediate and early-type galaxies, $\sigma_{S0}/\sigma_E$, (d) the 
concentration of the cluster C inside of $r_{200}$, (e) the ratio between 
the number of late- and early-type galaxies, $N_S/N_E$, and (f) the ratio 
between the number of intermediate- and early-type galaxies, $N_{S0}/N_E$.  
The upper axis, with small numbers, is labeled in redshift scale.  
The dotted and dashed lines in plots (a) and (d) correspond to the Coma and 
Virgo clusters, respectively.  The dot-and-dashed lines in plots (b) and (c) 
are the expected ratio of a falling over a virialized system, 
$\sigma_{fall}/\sigma_{vir}=\sqrt{2}$.  
The open diamonds correspond to clusters of the Dressler et al. 
(1997) sample.  \label{f5}}
\end{figure}

\begin{figure} 
\vspace{2cm}
\centerline{\epsfxsize= 15cm \epsfbox{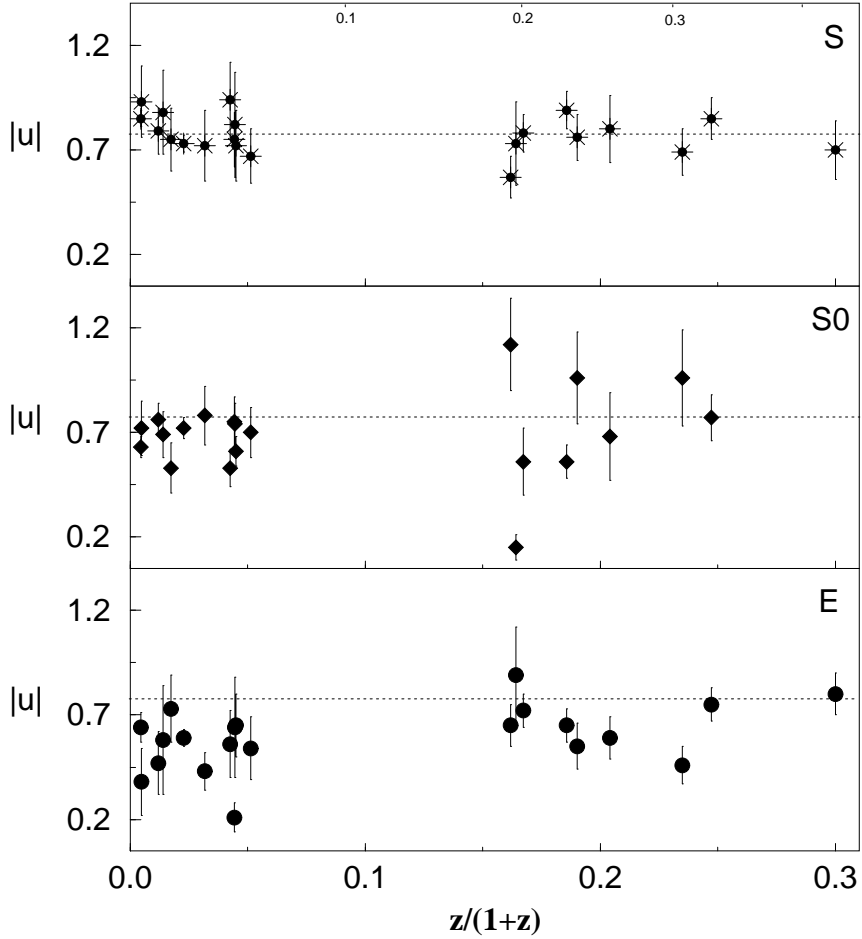}}
\caption{ 
Distribution of the velocity average deviation defined as $|u| =
<|v_i-v_{cl}|/\sigma_{cl}>$, where $v_{cl}$ and $\sigma_{cl}$ are the
mean velocity and velocity dispersion of the cluster calculated with
the bi-weighted estimator (Beers et al 1990). It is plotted as a
function of z/(1+z), which is proportional to the look-back time in a
universe with $\Omega_o$ = 0. The upper axis, with small numbers, is
labeled in redshift scale. The members were separated as 
late-type(upper plot), intermediate-type  (middle plot), or early-type 
(lower-plot) galaxies. Late-type galaxies, in the
clusters at intermediate-redshift are represented by galaxies with the
ratio of bulge to disk luminosity of B/T$\leq$0.4, intermediate-type 
galaxies by 0.4 $<$ B/T $<$ 0.7, and early-type galaxies by B/T$\geq$0.7.  
The dotted line represents the separation between 
eccentric orbits ($|u| < 0.77$) and approximately circular or
isotropic orbits ($|u| \geq 0.77$). As explained in section 2, the value 
$|u| = 0.77$ represents the highest value below which an eccentric orbit 
can be unambiguously identified. \label{f6} }
\end{figure}

\end{document}